# Traversal times for resonant tunneling

Uri Peskin

Department of Chemistry and the Lise Meitner Center for Computational Quantum Chemistry, Technion Israel Institute of Technology, Haifa 32000, Israel

And

Michael Galperin and Abraham Nitzan

School of Chemistry, the Sackler Faculty of Sciences, Tel Aviv University, Tel Aviv, 69978, Israel

## Abstract

The tunneling time of particle through given barrier is commonly defined in terms of "internal clocks" which effectively measure the interaction time with internal degrees of freedom of the barrier. It is known that this definition of the time scale for tunneling is not unique in the sense that it depends on the clock used to define it. For the case of resonance tunneling, a particular choice that in the limit of a high/broad square barrier yields the original result of Büttiker and Landauer (*Phys. Rev. Lett.* **1982**, *49*, 1739 ) is correlated to the lifetime of the resonance state. This is illustrated for analytically solvable one-dimensional double barrier models and for a realistic model of electron tunneling through a static water barrier. The latter calculation constitutes a novel application of this concept to a 3-dimensional model, and the observed structure in the energy dependence of the computed traversal time reflects the existence of transient tunneling resonances associated with instantaneous water structures. These models, characterized by the existence of shape resonances in the barrier, make it possible to examine different internal clocks that were proposed for measuring tunneling times in situations where a "clock independent" intrinsic time scale (the resonance life time) for the tunneling time exists. It is argued that this time may be used in order to estimate the relative importance of dynamical barrier processes that affect the tunneling probability.

# 1. Introduction

The dynamics of tunneling processes has been under discussion for a long time. 'Straightforward' timescales such as the rate for probability buildup on one side of a barrier following a collision of a particle wavepacket on the other side, or equivalently, the time associated with the tunneling splitting in a symmetric double well potential, are important measures of the *tunneling rate*. Following the work of Landauer and Büttiker[1-6] and others,[7] it has been recognized that other timescales may be relevant for other observables associated with the tunneling process. In particular, the question 'how long does the tunneling particle actually spends in the classically forbidden region of the potential' is of particular interest. This *traversal time for tunneling* is useful in estimates of the relative importance of processes that may potentially occur while the particle is in the tunneling region. Energy exchange with other degrees of freedom in the barrier and interaction with external fields focused in the barrier region (e.g. deflection of a tunneling electron by an electrostatic field induced by a heavy ion) are important examples.

The Büttiker-Landauer approach to tunneling timescales is based on imposing an internal clock on the tunneling system, for example a sinusoidal modulation of the barrier height.[1] At modulation frequencies much smaller than the inverse tunneling time the tunneling particle sees a static barrier which is lower or higher than the unperturbed barrier depending on the phase of the modulation. At frequencies much higher than the inverse tunneling time the system sees an average perturbation and so no effective change in the barrier height, but inelastic tunneling can occur by absorption or emission of modulation quanta. The inverse of the crossover frequency separating these regimes is the estimated traversal time for tunneling. For tunneling through the 1-dimensional rectangular barrier

$$V(y) = \begin{cases} U_B & ; \quad y_1 \leq y \leq y_2 \\ 0 & \quad otherwise \end{cases} \tag{1}$$

and provided that $d=y_2-y_1$ is not too small and that the tunneling energy $E$ is sufficiently below $U_B$, this analysis gives

$$\tau = \frac{d}{v_I} = \sqrt{\frac{m}{2(U_B - E)}} d \tag{2}$$

for a particle of mass *m* and energy $E < U_B$. $v_I$, defined by (2), is the imaginary velocity for the under-barrier motion. A similar result is obtained (see below) by using a clock based on population transfer between two internal states of the tunneling particle induced by a small barrier localized coupling between them.[2] Using the same clock for electron transfer via the super-exchange mechanism (equal-energy donor and acceptor levels coupled to opposite ends of a molecular bridge described by an N-state tight binding model with nearest-neighbor coupling V, with an energy gap $\Delta E$ V between the donor/acceptor and bridge states), yields[8]

$$\tau = \frac{\hbar N}{\Delta E} \tag{3}$$

It was shown[8] that both results (2) and (3) are limiting cases (wide and narrow band limits) of a more general expression.

The interpretation of $\tau$ defined above as a characteristic time for the tunneling process should be used with caution. An important observation made by Büttiker,[2] is that the tunneling time is not unique, and depends on the observable used as a clock. This observation is put on a more formal basis in section 2 below. Still, as shown in Ref. 1, it appears that with a proper choice of clock the traversal time provides a useful measure for the degree of adiabaticity of the interaction of the tunneling particle with barrier degrees of freedom. This issue is repeatedly encountered in electron tunneling through molecular environments, and its importance has been highlighted recently in studies of electron transport in metal-molecule-metal junctions. Whether the barrier appears rigid to the tunneling electron, as is often assumed in theoretical modeling, and to what extent inelastic transitions occur and affect transmission and conductance depend on the relative scales of barrier motions and traversal time, properly defined.

Assuming that Eqs. (2) and (3) do provide suitable measures for this purpose, we note that for a barrier of height $U_B - E \cong \Delta E \sim$ 1eV and width ~10Å (taking the corresponding *N* to be ~2-3), Eqs. (2) and (3) yield $\tau \cong$ 0.2fs and $\tau \cong$ 2fs, respectively, both considerably shorter than the period of molecular vibrations. When tunneling is affected or dominated by barrier resonances these estimates may change. For example, we have recently analyzed resonance effects in electron tunneling through water, and have shown that the lifetime of excess electron resonances associated with transient cavities in the water structure is of the order of 10fs, suggesting the possible involvement of OH vibrations and librations who move on similar timescales. A suitably defined tunneling time should contain similar information. Indeed, as shown in Ref. 9, the traversal time computed using the same clock that leads to Eqs. (2) and (3) shows a good correlation with the typical resonance lifetime. In the present paper we expand our analysis of this correspondence, using both a simple 1-dimensional double barrier model and a realistic 3-dimensional water barrier that corresponds to electron transfer between metal electrodes separated by a thin water film. A brief review of the clock-based definition of the tunneling traversal time is presented in Section 2. Section 3 discusses the applicability of the tunneling time concept to resonant tunneling processes within a simple 1-dimensional double barrier model, while Section 4 describes our results for water barriers. Section 5 concludes.

## 2. The traversal time as an observable variation problem

Different approaches to tunneling traversal and reflection times attempt to estimate the time the tunneling particle spends 'under the barrier' given that it is eventually transmitted or reflected. Additionally, a dwell time can be defined when the outcome of the tunneling process is undetermined.[10] These times are obviously different from each other. In the present section we focus on the traversal time. We follow the approach taken by Büttiker[2] following earlier works by Baz[11,12] and Rybachenko.[13] This approach is based on the analysis of a particle with two degenerate internal states undergoing elastic scattering (or tunneling) process. The two internal states are weakly coupled only in a particular spatial region (e.g. the barrier). An analysis of the subspace of internal states in different components of the outgoing wavefunction provides a measure of the time the particle spent in that region.

In the following discussion we consider a 1-dimensional tunneling process (defined to be along the *y* direction) and focus on the transmitted part of the outgoing wavefunction. The tunneling particle has an internal spin coordinate described in the space of two states, $|1>= \begin{pmatrix} 1 \\ 0 \end{pmatrix}$ and $|2>= \begin{pmatrix} 0 \\ 1 \end{pmatrix}$. These states are coupled only in the barrier region. The Hamiltonian describing the system is

$$H = \left[ -\frac{\hbar^2}{2m}\frac{\partial^2}{\partial y^2} + V(y) \right] \mathbf{I} + \lambda F(y)\sigma_x \tag{4}$$

where $F(y)$ is 1 in the barrier region and 0 outside it, **I** is the unit operator $\begin{pmatrix} 1 & 0 \\ 0 & 1 \end{pmatrix}$ and $\sigma_i$ are Pauli matrices, e.g. $\sigma_x = \begin{pmatrix} 0 & 1 \\ 1 & 0 \end{pmatrix}$. The Hamiltonian (4) (we use $H$ to denote the 2×2 Hamiltonian matrix) corresponds to the concrete example of a spin-1/2 particle described in the representation of eigenstates of the Pauli matrix $\sigma_z$ that interacts with a magnetic field $B$ pointing in the -$x$ direction, which vanishes outside the barrier and is constant inside it. In this case $\lambda = g\mu B/2 \equiv \hbar\omega_L/2$ where $g$ is the gyromagnetic ratio, $\mu$ is the absolute value of the magnetic moment and $\omega_L$ is the Larmor frequency.[14]

For an incident particle polarized in an arbitrary direction, i.e.,

$$\psi_{in} = \exp(iky)(a_1|1> + a_2|2>) = \exp(iky)\begin{pmatrix} a_1 \\ a_2 \end{pmatrix} \quad (5)$$

with $|a_1|^2 + |a_{21}|^2 = 1$, the transmitted wavefunction is, apart from the position dependent phase factor

$$\psi_{trans} = \begin{pmatrix} c_1 \\ c_2 \end{pmatrix} \quad (6)$$

where the incident and transmitted spinors are related by a linear transformation that depends on the barrier characteristics and on the coupling $\lambda$:

$$\mathbf{c} = \mathbf{S}\mathbf{a} \quad (7)$$

Analytical expressions for the elements of the scattering matrix S, as functions of the incident energy, for the Hamiltonian (4) for a 1-dimensional rectangular barrier were obtained by Büttiker.[2] In the limit $\lambda \to 0$ the total transmission probability $\mathcal{T} = |S_{11}|^2 = |S_{22}|^2$ is the same as would be obtained for a particle without internal structure. Association with time is achieved by considering the relationship between the normalized transmitted wavefunction, $\mathcal{T}^{-1/2}\psi_{trans}$, and the wavefunction obtained by the interaction $\lambda$ operating during time $\tau$, i.e.

$$\Psi(\tau) = e^{-i(H_\lambda/\hbar)\tau}\begin{pmatrix} a_1 \\ a_2 \end{pmatrix} \equiv \mathbf{U}(\tau)\mathbf{a} \quad (8)$$

where $H_\lambda = \lambda(|1><2| + |2><1|)$. For small $\lambda$

$$\mathbf{U}(\tau) = \begin{pmatrix} 1 - (1/2)(\lambda\tau/\hbar)^2 & -i\lambda\tau/\hbar \\ -i\lambda\tau/\hbar & 1 - (1/2)(\lambda\tau/\hbar)^2 \end{pmatrix} \quad (9)$$

We may attempt to define the traversal time for tunneling by formally requiring that for $\lambda \to 0$, $\Psi(\tau) = \mathcal{T}^{-1/2}\psi_{trans}$, i.e.[15]

$$\begin{aligned} U_{11}a_1 + U_{12}a_2 &= \mathcal{T}(\mathbf{a})^{-1/2}(S_{11}a_1 + S_{12}a_2) \\ U_{21}a_1 + U_{22}a_2 &= \mathcal{T}(\mathbf{a})^{-1/2}(S_{21}a_1 + S_{22}a_2) \end{aligned} \quad (10)$$

These equations, however, *are not mutually consistent*, i.e., give different results for $\tau$. Furthermore, these results may depend on the choice of the initial **a**. This is the origin of the observation that the 'tunneling time' depends on the observable used to estimate it. For a dynamical variable **A** in the spin sub-space, a time $\tau_A$ may be defined by the requirement that the relative change in **A** that accompanies the tunneling process is the same as that associated with the time evolution (8):

$$\frac{<\psi_{trans}|A|\psi_{trans}>}{<\psi_{trans}|\psi_{trans}>} - \frac{<\psi_{in}|A|\psi_{in}>}{<\psi_{in}|\psi_{in}>} = \frac{<\Psi(\tau_A)|A|\Psi(\tau_A)>}{<\Psi(\tau_A)|\Psi(\tau_A)>} - \frac{<\Psi(\tau=0)|A|\Psi(\tau=0)>}{<\Psi(\tau=0)|\Psi(\tau=0)>} \quad (11)$$

or

$$\mathcal{T}^{-1}\sum_{j=1}^{2}\sum_{k=1}^{2} A_{jk} c_j^* c_k = \sum_{j=1}^{2}\sum_{k=1}^{2} A_{jk} (\mathbf{U}(\tau_A)\cdot\mathbf{a})_j^* (\mathbf{U}(\tau_A)\cdot\mathbf{a})_k \quad (12)$$

Let the incident spin wave be in the +z direction, i.e., $\mathbf{a} = \begin{pmatrix}1\\0\end{pmatrix}$. For $\mathbf{A}=\sigma_z=\begin{pmatrix}1 & 0\\0 & -1\end{pmatrix}$ this leads to

$$\frac{|S_{11}|^2 - |S_{21}|^2}{|S_{11}|^2 + |S_{21}|^2} = 1 - 2\left(\frac{\lambda\tau_z}{\hbar}\right)^2, \quad (13)$$

to second order in $\lambda$, while $\mathbf{A}=\sigma_y=\begin{pmatrix}0 & -i\\i & 0\end{pmatrix}$ yields

$$\frac{\text{Im}(S_{11}S_{21})}{|S_{11}|^2 + |S_{21}|^2} = \frac{\lambda\tau_y}{\hbar} \quad (14)$$

These equalities yield expressions for $\tau_z$ and $\tau_y$ which are consistent with Büttiker's definitions[2] of the corresponding tunneling times in terms of the spin rotations in the two directions which are orthogonal to the direction of the "external" magnetic field.[16] However, using $\mathbf{A}=\sigma_x=\begin{pmatrix}0 & 1\\1 & 0\end{pmatrix}$ gives zero on the r.h.s of (12), so no information can be obtained on the tunneling time $\tau_x$, which is associated by Büttiker with spin rotation in the direction parallel to the external filed. Indeed, the commutativity of $\sigma_x$ with $H$ implies that $<\sigma_x>$ remains zero at all time under the time evolution (8).[17] We note that Büttiker[2] has identified $<\sigma_x>$ with $\omega_L \tau_x$, and *defined* $\tau_x$ by this relation. This leads to

$$\frac{\text{Re}(S_{11} S_{21}^*)}{|S_{11}|^2 + |S_{21}|^2} = \frac{\lambda \tau_x}{\hbar} \qquad (15)$$

These three times are related by[2]

$$\tau_x^2 + \tau_y^2 = \tau_z^2$$

The fact that the tunneling time obtained as described above depends on the observable used to define it is an awkward feature of this concept. It is interesting to note that, for an incident particle in the $\mathbf{a} = \begin{pmatrix} 1 \\ 0 \end{pmatrix}$ state, if we replace the requirement $\Psi(\tau) = \mathcal{T}^{-1/2} \psi_{trans}$, i.e. $\mathbf{U} = \mathcal{T}^{-1/2} \mathbf{S}$ by $|U_{ij}| = \mathcal{T}^{-1/2} |S|_{ij}$ (to be evaluated for $\lambda \to 0$) it may be easily checked that the resulting two equalities lead both to the same expression for $\tau$,

$$\tau = \frac{\hbar}{|\lambda|} \frac{|S_{12}|}{\mathcal{T}^{1/2}} = \frac{\hbar}{|\lambda|} \frac{|S_{12}|}{|S_{11}|} = \tau_z \qquad (16)$$

which, for a square barrier characterized by the width $d$ and height $U_B$, yields the result (2).[2]

This definition of the tunneling time $\tau$ as $\tau_z$ is appealing as a measure of the duration for a tunneling process for the purpose of considering the importance of competing population transfer processes in the barrier. Furthermore, in the following sections we find (see also Ref. 18) that for resonance tunneling this time correlates well with the lifetime of the barrier resonance. Still, it should be emphasized that this concept should be used with caution. For example tunneling times defined by Eqs. (10) or (12) also depend on the energy spacing between the two internal (spin) levels that was taken zero above. For example, Fig. 1 shows the dependence of the time $\tau$ of Eq. (16) on the energy spacing $\Delta E = E_1 - E_2$ between the levels $|1\rangle$ and $|2\rangle$, i.e. the Hamiltonian (4) is replaced by

$$H = \left[ -\frac{\hbar^2}{2m} \frac{\partial^2}{\partial y^2} + V(y) \right] \mathbf{I} + \frac{1}{2} \Delta E \sigma_z + \lambda F(y) \sigma_x \qquad (17)$$

It is seen that the computed traversal time depends on $\Delta E$, not a surprising result considering the fact that at constant incident energy the two internal states see different effective barriers that depend on $\Delta E$. Still, in the range $-0.5\text{eV} < \Delta E < 0.5\text{eV}$, which is the relevant range for assessing the relative importance of nuclear dynamics effects on electron tunneling, this dependence is seen to be modest and the calculated time provides a reasonable order of magnitude indication.

**3. Traversal times in resonant tunneling by the distorted wave approach**

As argued above, the concept of tunneling time is useful when discussing the possible importance of barrier processes that transfer population between internal states of the tunneling system. Resonant tunneling situations are important examples of cases were such barrier processes could happen. Here we apply the formalism outlined above to such processes. A 1-dimensional double barrier model where the tunneling behavior is affected by resonance(s) in the intermediate well is considered in this section, and a 3-dimensional model that corresponds to a water layer between two metal electrodes is discussed in Sect. 4. In both cases we show that near resonance the result of Eq. (16) is in a good correspondence with the resonance lifetimes, while alternative measures can give counter-intuitive results.

The one dimensional double barrier potential is defined in terms of the potential energy function:

$$V(y) = \begin{cases} 0 & y > y_4 \\ V_0 & y_3 < y \le y_4 \\ 0 & y_2 < y \le y_3 \\ V_0 & y_1 < y \le y_2 \\ 0 & y \le y_1 \end{cases} \quad (18)$$

The tunneling time through the double barrier is defined in terms of the two dimensional Hamiltonian in Eq. (4), in which the one dimensional tunneling coordinate is coupled locally to internal "spin" levels. The coupling range is defined by $F(y)$, which is zero for $y<y_1$ and $y>y_4$, and equals 1 within the barrier range, $y_1<y<y_4$. In the particular case of resonant tunneling through a symmetric double barrier, where the on-resonance transmission probability is unity, it was shown by Leavens and Aers[18] that $\tau_x$ vanishes, and so the traversal time is given by

$$\tau_{trav} = \tau_z = \tau_y. \quad (19)$$

In the previous section the different tunneling times were defined in terms of the scattering matrix elements associated with the two dimensional ("spinor") Hamiltonian, Eq.(4). Our purpose is to express $\tau_{trav}$ in terms of the parameters of the one-dimensional double barrier (Eq.(18)). We start by dividing the potential energy operator into two parts, following the two potential formalism,[19]

$$V = V_I + V_{II} \quad (20)$$

with

$$V_I = V(y)\mathbf{I}_2 \quad ; \quad V_{II} = \lambda F(y)\sigma_x \quad (21)$$

In the weak coupling limit, $\lambda \to 0$, the transmission probability amplitudes can be well approximated by the distorted waves approximation. Let the incoming wave vector be $k = \sqrt{2mE/\hbar^2}$, the transmission probability amplitude from an initial spinor state $|\varphi_j\rangle\rangle$ to a final spinor state $|\varphi_i\rangle\rangle$ is given by,

$$S_{i,j} = \delta_{i,j} - \frac{im}{k\hbar^2} <<\varphi_i | V | \psi_j>> \xrightarrow{\lambda \to 0}$$
$$\delta_{i,j} - \frac{im}{k\hbar^2}(<<\psi_{I,i}^- | V_{II} | \psi_{I,j}^+ >> + <<\psi_{I,i}^- | V_I | \varphi_j>>) \quad (22)$$

The double bracket notation is used for integration over both the spatial and the spin coordinates. $|\psi_j>>$ is the exact scattering wave function associated with the asymptotic state $|\varphi_j>>$

$$|\psi_j>> = (1 + \frac{1}{E - H + i\varepsilon} V) |\varphi_j>> \quad (23)$$

and the asymptotic states are defined as $|\varphi_1>> = e^{iky}\begin{pmatrix}1\\0\end{pmatrix}$ and $|\varphi_2>> = e^{iky}\begin{pmatrix}0\\1\end{pmatrix}$. $|\psi_{I,i}^{\pm}>>$ are the distorted waves, which are the exact incoming (+) and outgoing (-) scattering states in the absence of the two-level coupling (i.e., for $\lambda = 0$):

$$|\psi_{I,j}^{\pm}>> = (1 + \frac{1}{E - H_I \pm i\varepsilon} V_I) |\varphi_j>> \quad (24)$$

Substitution of the appropriate asymptotic states leads to the following expression for the transmission probability amplitudes:

$$S_{1,1} = 1 - \frac{im}{k\hbar^2} <<\psi_{I,1}^- | V_I | \varphi_1>> = 1 - \frac{im}{k\hbar^2} \int_{y_1}^{y_4} \psi_{I,1}^{-*}(y) V(y) \varphi_1(y) dy \quad (25)$$

$$S_{1,2} = -\frac{im}{k\hbar^2} <<\psi_{I,1}^- | V_{II} | \psi_{I,2}^+>> = -\frac{im\lambda}{k\hbar^2} \int_{y_1}^{y_4} \psi_{I,1}^{-*}(y) \psi_{I,1}^+(y) dy \quad (26)$$

Using the equations (13), (25), (26), the tunneling time $\tau_z$ can be expressed as:

$$\tau_z = \hbar \sqrt{\frac{|S_{1,2}|^2}{\lambda^2(|S_{1,1}|^2 + |S_{1,2}|^2)}} \approx \frac{\hbar |S_{1,2}|}{\lambda |S_{1,1}|} = \frac{m}{k\hbar |S_{1,1}|} | \int_{y_1}^{y_4} \psi_{I,1}^{-*}(y) \psi_{I,1}^+(y) dy | \quad (27)$$

where we have used the limit $\lambda \to 0$ again, in neglecting $|S_{1,2}|^2$ relative to $|S_{1,1}|^2$. The latter expression can be farther simplified for the particular case of a *symmetric* double barrier, with isolated resonance states. For $\lambda \to 0$ the transmission at near resonance impact energies is close to unity, $|S_{1,1}| \to 1$. In such a case, one obtains

$$\tau_z = \frac{X_{eff}}{\upsilon} \tag{28}$$

where $\upsilon$ is the incoming velocity

$$\upsilon = \frac{k\hbar}{m} \tag{29}$$

and the effective barrier width, $X_{eff}$, is given by the overlap integral between the incoming and outgoing distorted waves in the barrier region:

$$X_{eff} = |\int_{y_1}^{y_4} \psi_{I,1}^{-*}(y)\psi_{I,1}^{+}(y)dy| \tag{30}$$

The distorted waves can be calculated explicitly as solutions to the one-dimensional Schrödinger equation,

$$\left[\frac{-\hbar^2}{2m}\frac{\partial^2}{\partial y^2} + V(y)\right]\psi(y) = E\psi(y) \tag{31}$$

For a stepwise potential,

$$V(y) = V_n \quad ; \quad y_n \leq y < y_{n+1} \quad ; \quad n = 1,2,...,N \tag{32}$$

a solution to the Eq. (31) in the $n^{th}$ segment is given by

$$\psi_n(y) = A_n(E)e^{ik_n y} + B_n(E)e^{-ik_n y} \quad ; \quad k_n = \sqrt{\frac{2m(E-V_n)}{\hbar^2}} \tag{33}$$

where the standard continuity condition of the function and its derivative at the matching points leads to the following recursion[20]:

$$\frac{A_{n-1}(E)}{B_{n-1}(E)} = e^{-2ik_{n-1}y_n}\frac{(A_n(E)/B_n(E))e^{ik_n y_n}(k_{n-1}+k_n) + e^{-ik_n y_n}(k_{n-1}-k_n)}{(A_n(E)/B_n(E))e^{ik_n y_n}(k_{n-1}-k_n) + e^{-ik_n y_n}(k_{n-1}+k_n)} \tag{34}$$

The coefficients $\{A_n(E)\}$ and $\{B_n(E)\}$ of the incoming and outgoing distorted waves are obtained by the recursion above, with the following incoming wave boundary conditions,

$$\begin{aligned}\psi_{I,1}^{+}(y) &\leftrightarrow \quad A_1 = 1; \quad B_N = 0 \\ \psi_{I,1}^{-}(y) &\leftrightarrow \quad A_N = 1; \quad B_1 = 0\end{aligned} \tag{35}$$

The tunneling time, $\tau_z$, can therefore be calculated explicitly according to Eqs.(28-30).

Let us consider two different double barrier model potentials. The first is a symmetric double barrier potential characterized by the parameters $U_B$=5eV, $(y_1, y_2, y_3, y_4)$ = (-10, -7.5, 7.5, 10) au. The second is non symmetric with the parameters $U_B$=5eV, $(y_1, y_2, y_3, y_4) = (-10, -7.5, 3.75, 10)$ au. Both models support resonance states, which are solutions of the Schrödinger equation (which amounts to the recursion relation, Eq.(34)) when outgoing wave boundary conditions are applied. Denoting the resonance state as $\psi_R(y)$, the corresponding boundary conditions for the recursion are,

$$\psi_R(y) \quad \leftrightarrow \quad A_1 = 0; \quad B_N = 0 \qquad (36)$$

These conditions can only be satisfied for a complex resonance energy,

$$E = E_r - i\frac{\Gamma}{2} \qquad (37)$$

A search was carried out for complex resonance energies for the two models. The resonance lifetimes, $\tau_{res} = 1/\Gamma$, were calculated directly from the imaginary parts of the resonance energies and are given in Table I. For each resonance the traversal time, $\tau_{trav}$, was calculated in terms of the distorted waves (Eqs. (19) and (27)), for the impact energy $E=E_r$, the real part of the resonance energy. These results are also shown in Table I. We see that the resonance lifetimes and the traversal times are strongly correlated. Interestingly, we find that "on resonance" the traversal time which measures the interaction time with an external clock is roughly twice the lifetime of the corresponding resonance states. One may be tempted to explain this observation for a symmetric double barrier where the resonance decays to both directions at double the rate associated with traversal to one direction. Indeed, for this case Büttiker[21] has shown that within the Breit-Wigner approximation for the scattering matrix near resonance, the tunneling time is equal to the dwell time, which is twice the resonance lifetime.

| Model | Resonance Energy | $\tau_{res}$ | $\tau_{trav}$ | $\tau_{trav}/\tau_{res}$ |
|---|---|---|---|---|
| Symmetric | 0.0142 - i0.000302 | 1655.6 | 3330 | 2.0114 |
|  | 0.0566 - i0.002517 | 168. | 410 | 2.0639 |
|  | 0.1266 - i0.008990 | 55.617 | 122 | 2.1936 |
| Non symmetric | 0.02265 - i0.0003915 | 1277.1 | 2574 | 2.0155 |
|  | 0.0887 - i0.0032165 | 155.45 | 324 | 2.0843 |

**Table I**: Resonance lifetimes and traversal times for double barrier potentials.

Interestingly, we obtain similar results also for a non-symmetric barrier, which suggests that the roots of this observation may run deeper and will be addressed elsewhere.

## 4. Traversal times in water

In this section we examine the use of the traversal time concept for a particular and singularly important case - electron tunneling through water. Preliminary results for this system were already published.[9] For specificity we consider a particular situation: electron tunneling through a water layer confined between two planar Pt (100) electrodes. Our model system and interaction potentials are the same as those used before[22,23] to evaluate electron transmission probabilities in water. In particular, the potential experienced by the electron is taken to be a superposition of the vacuum potential, modeled by a rectangular barrier, and the electron-water interaction.[24] The latter is represented by the pseudo-potential of Barnett et al,[25] modified[26] to account for the many-body aspect of the water electronic polarizability. Water configurations are sampled from an equilibrium trajectory obtained by running classical molecular dynamics simulations. The electron Hamiltonian is represented on a grid in position space. The overall grid size that was used is 16×400×16, with grid spacings 0.4au in the tunneling direction (*y*) and 2.77au in the parallel directions (*x*,*z*). Absorbing potentials, applied near the grid boundary in the *y* direction, make it possible to solve a scattering problem on a finite grid. Periodic boundary conditions are used in the *x* and *z* directions. The distance between the metal electrodes depends on the number of water monolayers. The overall dimensions of the water slab in the simulation cell were thus 23.5×10×23.5Å for 3 monolayers, and 23.5×12.9×23.5Å for 4 monolayers. The water density between the electrodes was assumed independent of the confinement, and was taken 1 g/cm$^3$. This corresponds to a of total 197 and 257 water molecules in these two water slabs.

We consider the *one-to-all* transmission probability: the electron is incident in the direction *y* normal to the barrier, and the transmission probability is a sum over all final directions. For an electron without internal states, described by a Hamiltonian $H_0$, this probability is given by[27,28]

$$T = \frac{2}{\hbar} < \phi_{in}(E) | \varepsilon_{in} G \ \varepsilon_{out} G \varepsilon_{in} | \phi_{in}(E) > \qquad (38)$$

where $\phi_{in} = e^{ikx}/\sqrt{\upsilon}$ with $k = \sqrt{2mE/\hbar^2}$ and $\upsilon = \hbar k/m$, $\varepsilon_{in}$ and $\varepsilon_{out}$ are absorbing potentials in the incident and transmitted wave regions and

$$G = (E - H_0 + i(\varepsilon_{in} + \varepsilon_{out}))^{-1} \qquad (39)$$

For the absorbing potentials we have used $\varepsilon_{in} = (2|y|/L_y)^7$ for $-L_y/2 < y < 0$ and the corresponding mirror image for $\varepsilon_{out}$, where $L_y$ is the length of the calculational grid in the $y$ direction.

For the present problem we take $|\phi_{in}\rangle\rangle = (e^{ikx}/\sqrt{\upsilon})\begin{pmatrix}1\\0\end{pmatrix}$ and the Green's operator is given by (39) with $H_0$ replaced by

$$H = H_0 \begin{pmatrix} 1 & 0 \\ 0 & 1 \end{pmatrix} + \lambda F(y) \begin{pmatrix} 0 & 1 \\ 1 & 0 \end{pmatrix} \qquad (40)$$

where $\lambda$ is a constant and $F(y)=1$ in the barrier region and 0 outside it. The approximate scattering wave function,

$$|\psi(E)\rangle\rangle = iG(E)\varepsilon_{in}|\phi_{in}(E)\rangle\rangle = \begin{pmatrix}\psi_1(E)\\\psi_2(E)\end{pmatrix} \qquad (41)$$

is evaluated using iterative inversion methods as in our previous work.[22,23] The transmission probabilities into the |1> and |2> states (summed over all final directions of the transmitted wave) are obtained from[27,28]

$$|S_{i1}(E)|^2 = (2/\hbar)\langle\langle\psi_i(E)|\varepsilon_{out}|\psi_i(E)\rangle\rangle \quad ; \qquad i = 1,2 \qquad (42)$$

These probabilities are used to compute the traversal time $\tau(E)$ according to Eq. (16)

Figure 2 shows calculated traversal times as functions of incident electron energy. The distance between the two platinum electrodes is here $d$=18.9au, corresponding to three water monolayers. The barrier potential is taken as the superposition of the vacuum potential (represented by a simple rectangular barrier of height $U_B$) and the electron-water effective potential. Shown are the results obtained for this barrier (full line) and for the corresponding vacuum potential (dashed line). The dotted line represents the approximation (2) to the traversal time for the vacuum potential. These results were obtained for a vacuum barrier height $U_B$=5eV, but taking $U_B$=3eV made practically no difference. We may conclude that, as in Eq. (2), also for the 3-dimensional water barrier the traversal time depends mainly on the incident energy measured relative to the (vacuum) barrier height and only very weakly on the absolute energy. Two other significant observations can be made: (a) For the 3-dimensional water barrier the tunneling time exhibits a complex dependence on the incident energy (measured relative to the vacuum energy), and in particular what appear to be resonance features are seen below the vacuum barrier. (b) The absolute traversal times are fractions of fs in the deep tunneling regime, and 5-10fs at the peaks of the resonance structure below the vacuum barrier.

It should be emphasized that the results displayed in Fig. 2 correspond to a single static configuration of the equilibrated water, and different results, in particular in the resonance-dominated tunneling regime within 1eV below the vacuum barrier, are obtained for different static configurations. The following common features are noteworthy.[9] First, a strong correlation with the resonance structure of the transition probability is observed (see also below); secondly, in the deep tunneling regime, the computed time is similar for different configurations, is proportional to the barrier width and is ~10% longer than for the vacuum barrier.

The nature of the resonance structure observed below the vacuum barrier is elucidated in Figure 3. Here we show, for a particular configuration of the 3-monolayer film, the tunneling time and the transmission probability, both as functions of the incident electron energy. The resonance structure in the transmission probability was discussed in Ref. 23 and was shown to be associated with cavities in the water structure. Here we see that the energy dependence of the tunneling time follows this resonance structure closely. In fact, the times (3-15 fs) obtained from the peaks in Figs. 3 correlate well with the resonance lifetimes estimated in Ref. 23. A similar correspondence was found for all configurations studied. It is interesting to note that, as in the 1-dimensional case discussed above, the traversal times at resonance energies are longer by factors of order 2 than the corresponding resonance lifetimes. For example, for several of the resonances found in Ref. 23 we find an average traversal 11fs, compared with an average lifetime of 6.65fs. We leave the exact correspondence between these times to future study.

The calculations discussed so far are based on the one-to-all transmission probability with the electron incident normal to the water layer. For completeness we consider also the equivalent result obtained from a one-to-one transmission probability, where the electron is transmitted at a given angle relative to the layer. The needed $S$ matrix elements were calculated from Eq. (41) using[27-29]

$$S_{i1} = <<\phi_i | \varepsilon_{out} | \psi >> \tag{43}$$

with $|\psi>>$ given by Eq. (41) $|\phi_1>> = \upsilon^{-1/2} e^{i\mathbf{k}\cdot\mathbf{r}} \begin{pmatrix} 1 \\ 0 \end{pmatrix}$; $|\phi_2>= \upsilon^{-1/2} e^{i\mathbf{k}\cdot\mathbf{r}} \begin{pmatrix} 0 \\ 1 \end{pmatrix}$.

Fig. 4 shows the results obtained for $\tau$ (our $\tau_z$; $\tau_x$ of Ref. 2) as well as $\tau_x$ and $\tau_y$ (Eqs. (15) and (14); $\tau_z$ and $\tau_y$ of Ref. 2). We see that the estimate for $\tau$ is only weakly sensitive to the 'experiment' (one-to-one or one-to-all) done. Also, as functions of incident energy, $\tau_x$ and $\tau_y$ behave quite differently from $\tau$. In particular, $\tau_x$ shows a pronounced dip (familiar from earlier studies[18]) near the resonance energy, while $\tau_y$ can become negative close to the barrier top where interference features affect the transmission probability. Surprisingly, we find that these times are not very sensitive to the particular incident and scattered directions used in the 1-to-1 calculation.

These calculations were carried using static water structures sampled from a classical equilibrium distribution. The computed times provide a posteriori justification for this procedure. In particular, the relatively long times obtained near the resonance peaks are short relative to the lifetime of the structural defects that give rise to these resonances. It is important to note, however, that these times are of the same order of magnitude as the periods of intermolecular librations and intramolecular OH stretch vibrations, suggesting the possibility that inelastic processes contribute to the tunneling process. Indeed, recent calculations[30] have elucidated the effects of water vibrational and librational motions on electron tunneling through this system.

## 5. Conclusions

The tunneling time is not a unique quantity, and different measures depend on the observables used to quantify them. We have pointed out that a particular measure, Eq. (16), is particularly appealing in that at resonance it is closely correlated with the resonance lifetime computed independently. We have applied this measure in order to compute this tunneling time for electron traversing a water barrier separating two metal electrodes. As in 1-dimensional rectangular barrier model, in the deep tunneling regime (>1eV below the barrier) the computed time was found to depend mainly on the relative energy barrier rather than on the absolute incident electron energy, and to be proportional to the distance between the electrodes. For distances of the order of ~10Å the computed times in this regime are in the range of 0.1-1fs. Within 1eV from the vacuum barrier a marked structure in the energy dependence of the tunneling time is associated with resonances originating from structural defects in the water structure.[23] The tunneling times, ~10fs, computed at the peaks of these structures, follow the lifetimes of the corresponding resonances. These results set the scale for gauging possible effects of other barrier motions, e.g., intramolecular water vibrational modes, on the tunneling process.

**Acknowledgements**. This work is dedicated to Professor John Tully, a friend, a scientist and a source of inspiration on many aspects of quantum dynamics. This research was supported by the U.S-Israel Binational Science Foundation (AN), and by the Israel Science Foundation and by the fund for promotion of research at the Technion (UP).

**Figure Captions**

Fig. 1. The traversal time, Eq. (16), for a system described by the Hamiltonian (17) with $V(y)$ given by Eq. (1), displayed as a function of $\Delta E$ for a particle incident on a 1-dimensional barrier with energy $E_0$ in channel 1, i.e. in spin state $|1\rangle = \begin{pmatrix} 1 \\ 0 \end{pmatrix}$. The barrier height is $U_B = 5\text{eV}$ and the incident kinetic energy is 2eV (full line), 3eV (dashed line) and 4eV (dotted line).

Fig.2  The computed traversal time as a function of the incident electron energy measured relative to the vacuum barrier. See text for details.

Fig. 3. The traversal time $\tau$ (full line; left vertical scale) and the transmission probability (dotted line; right vertical scale) computed as functions of incident electron energy for one static configuration of the 3-monolayer water film.

Fig. 4. Different "tunneling times" vs. incident electron energy for the configuration of Fig. 3. All results are for incident direction normal to the water layer. (a) The traversal time $\tau$ obtained from the one-to-all transmission (full line, same as full line of Fig. 3), and for one-to-one transmission with outgoing wave at 20 degrees (dashed line) and 45 degrees (thin dotted line) to the normal. (b) One-to-one transmission calculation of $\tau_y$ (full line and thick-dotted line - outgoing wave at 20 and 45 degrees to normal, respectively) and of $\tau_x$ ($\tau_z$ of Ref. 2; dashed line and thin-dotted lines - outgoing wave at 20 and 45 degrees to normal, respectively)

# References

1. M. Büttiker and R. Landauer, *Phys. Rev. Lett.* **49,** 1739-1742 (1982).
2. M. Büttiker, *Phys. Rev. B* **27,** 6178-6188 (1983).
3. M. Büttiker and R. Landauer, *Physica Scripta* **32,** 429 (1985).
4. T. Martin and R. Landauer, *Phys. Rev. A* **47,** 2023 (1993).
5. R. Landauer and T. Martin, *Rev. Mod. Phys.* **66,** 217-228 (1994).
6. R. Landauer, 95, 404, *Ber. Bunsenges Phys. Chem.* **95,** 404 (1991).
7. E. H. Hauge and J. A. Stoveng, *Rev. Mod. Phys.* **61,** 917-936 (1989).
8. A. Nitzan, J. Jortner, J. Wilkie, A. L. Burin, and M. A. Ratner, *J. Phys. Chem. B, 104, 5661-5665 (2000)* **104,** 5661-5665 (2000).
9. M. Galperin, A. Nitzan, and U. Peskin, *to be published*, (2001).
10. See Ref. (5) and references therein.
11. A. I. Baz, *Sov. J. Nucl. Phys.* **4,** 182-188 (1967).
12. A. I. Baz, *Sov. J. Nucl. Phys.* **5,** 161-164 (1967).
13. V. F. Rybachenko, *Sov. J. Nucl. Phys.* **5,** 635-639 (1967).
14. Note that the formulation in terms of Pauli spin matrices uses common notations $\sigma_x, \sigma_y, \sigma_z$ for these matrices, however for particular applications like the concrete example discussed here, the directions implied by the spin subscripts x,y,z are not necessarily related to the physical directions, that where taken so that the tunneling is in the y direction. Also note that in Ref. 2 the Hamiltonian (4) is written in a diagonal representation $H = \left[ -\left( \hbar^2/2m \right) \partial^2/\partial y^2 + V(y) \right] \mathbf{I} - \lambda F(y) \boldsymbol{\sigma}_z$, obtained from Eq. (4) using a unitary transformation affected by $\mathbf{U} = \frac{1}{\sqrt{2}} \begin{pmatrix} 1 & 1 \\ -1 & 1 \end{pmatrix}$, i.e. $\mathbf{U}^\dagger \mathbf{U} = 1$ and $\mathbf{U}^\dagger \boldsymbol{\sigma}_z \mathbf{U} = -\boldsymbol{\sigma}_x$.
15. Obviously, equivalent definition can be made for the reflection time.[2]
16. See Ref. 2. Note that in our notation the role of the x and z spin directions are interchanged relative to this reference.[14]
17. The l.h.s. of (12) does not vanish for $\mathbf{A}=\sigma_x$ because the transmitted wavefunction $\psi_{trans}$ does not result from a time evolution of $\psi_{in}$. The latter evolve into a combination of a transmitted and reflected waves, for which it may be shown that indeed $<\sigma_x>$ does not change from its value in the incident state.
18. C. R. Leavens and G. C. Aers, *Phys. Rev. B* **40,** 5387-5400 (1989).

[19] J. R. Taylor, *Scattering Theory* (Wiley, New York, 1972).

[20] U. Peskin, *J. Chem. Phys.* **113,** 7479 (2000).

[21] M. Büttiker, in *Electronic Properties of Multilayers and low Dimensional Semiconductors*, Vol. 231, edited by J. M. Chamberlain, L. Eaves, and J. C. Portal (Plenum Press, New York, 1990).

[22] A. Nitzan and I. Benjamin, *Accounts of Chemical Research* **32,** 854-861 (1999).

[23] U. Peskin, A. Edlund, I. Bar-On, M. Galperin, and A. Nitzan, *J. Chem. Phys.* **111,** 7558-7566 (1999).

[24] This is obviously a simplified model. The separability of these contributions is only an approximation, and the vacuum potential is distorted because of image forces.

[25] R. N. Barnett, U. Landmann, C. L. Cleveland, and J. Jortner, *J. Chem. Phys.* **88,** 4421-4428 (1988).

[26] A. Mosyak, P. Graf, I. Benjamin, and A. Nitzan, *Journal of Physical Chemistry a* **101,** 429-433 (1997).

[27] T. Seideman and W. H. Miller, *J. Chem. Phys* **97,** 2499 (1992).

[28] T. Seideman and W. H. Miller, *J. Chem. Phys* **96,** 4412 (1992).

[29] W. H. Thompson and W. H. Miller, *Chem. Phys. Lett.* **206,** 123 (1993).

[30] M. Galperin and A. Nitzan, *J. Chem. Phys.* **115,** 2681-2694 (2001).

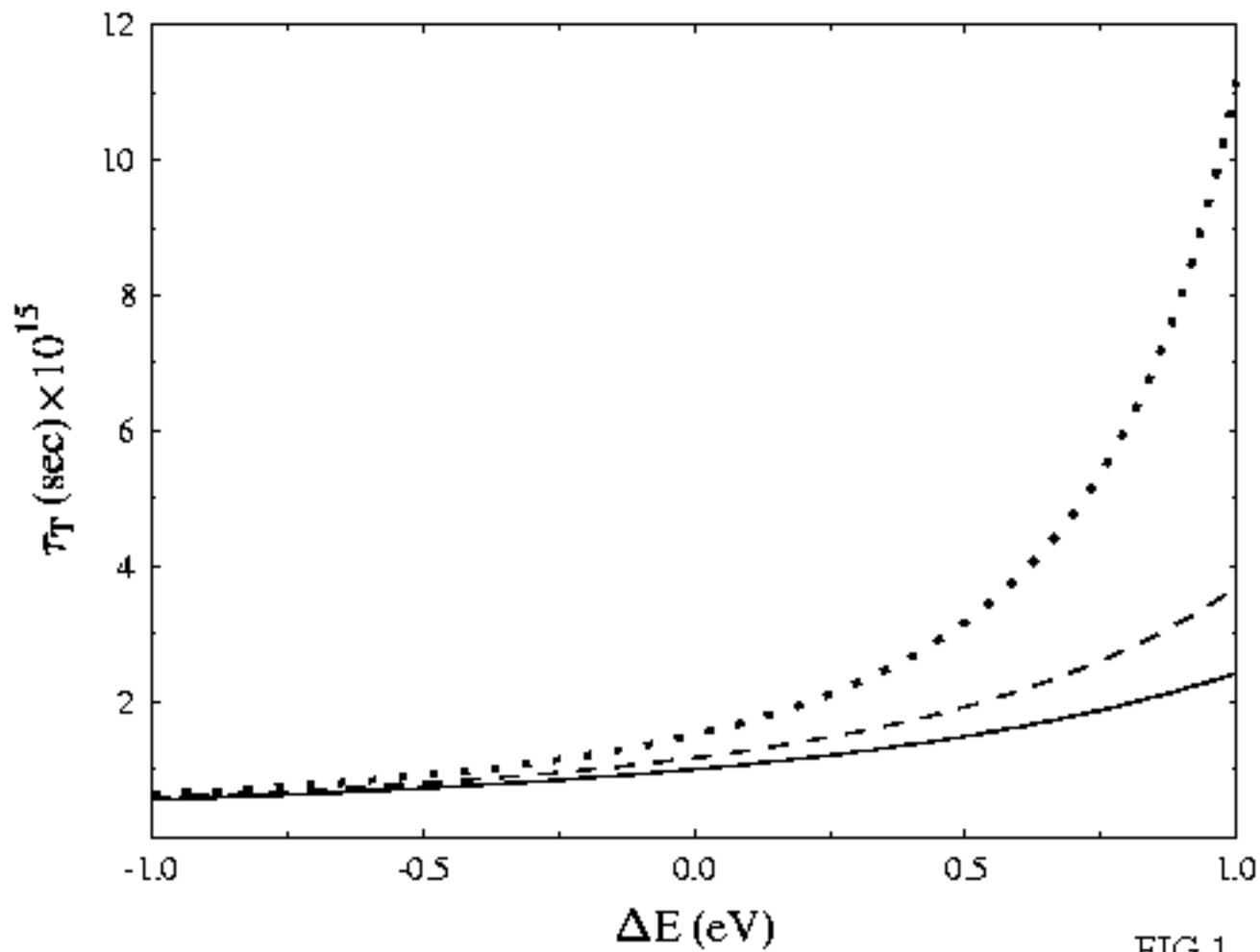

FIG.1

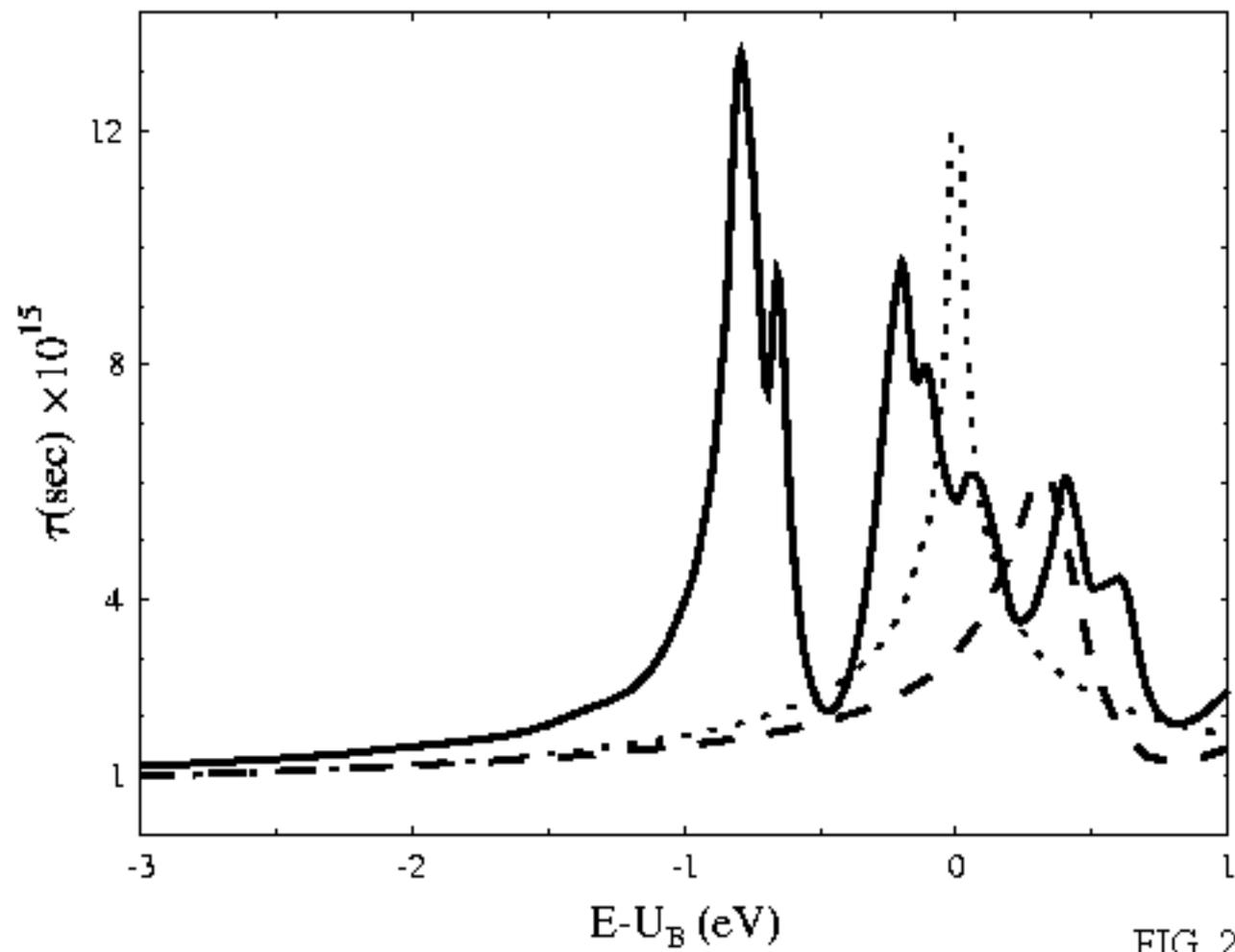

FIG. 2

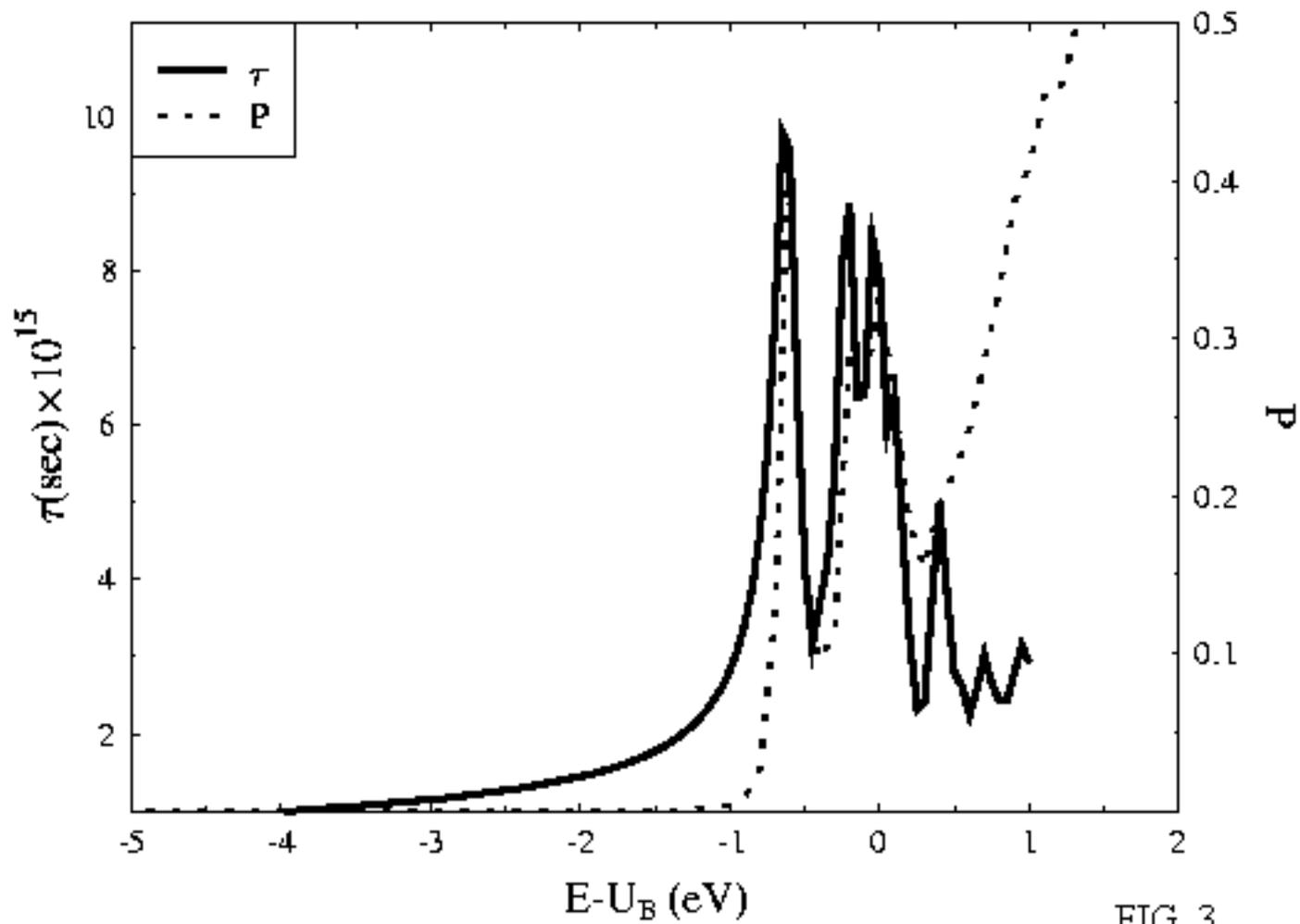

FIG. 3

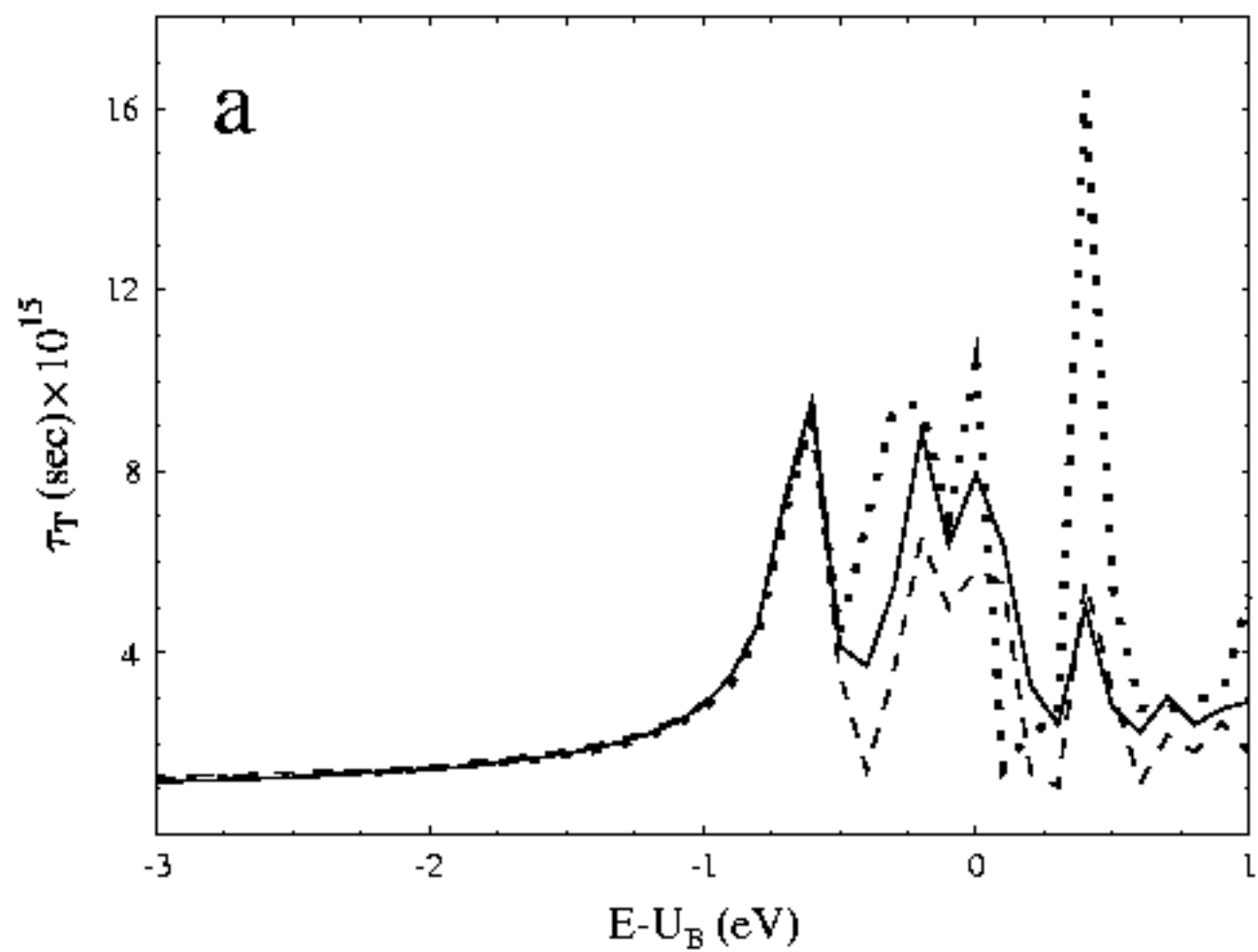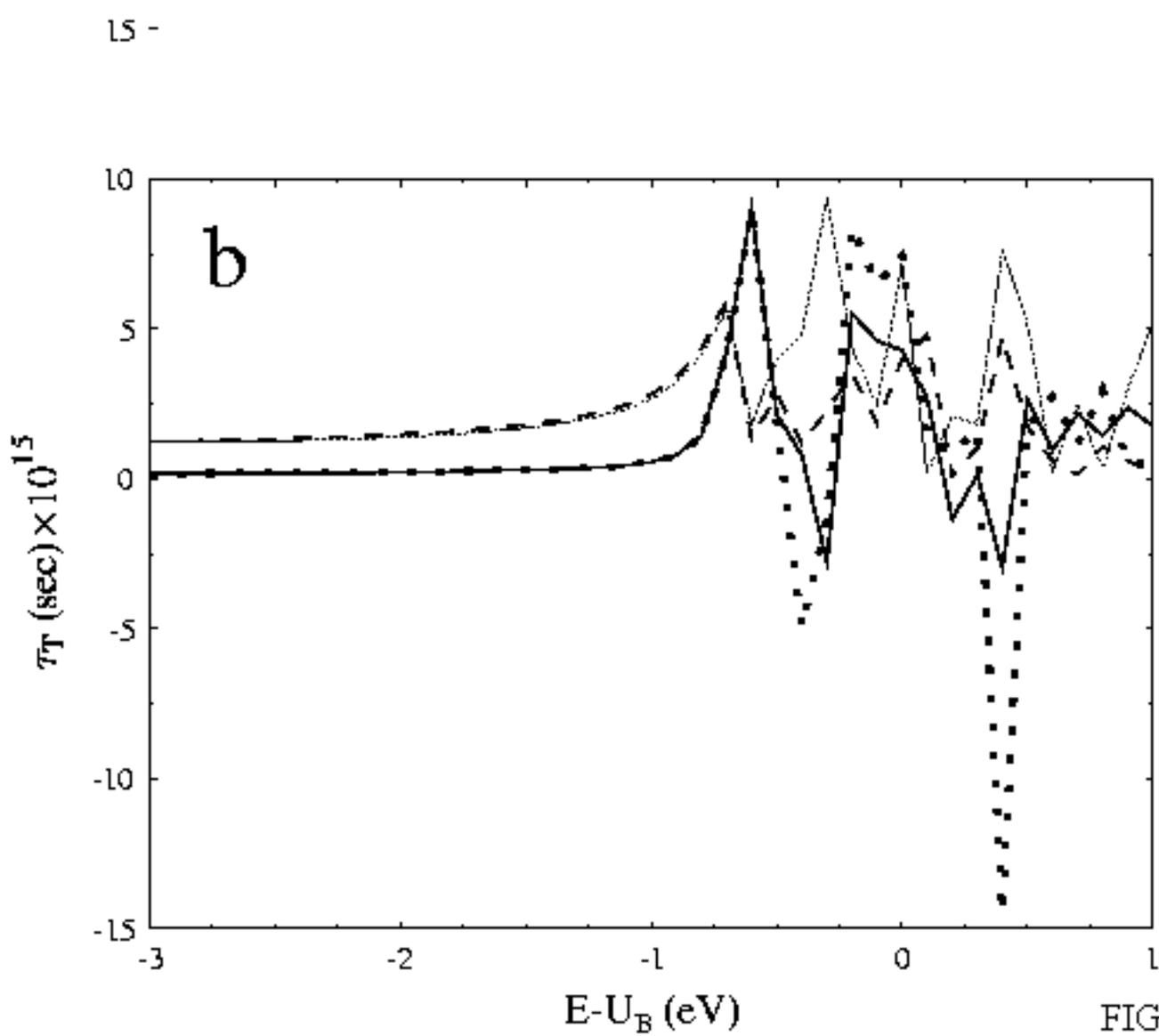

FIG. 4